\documentclass[english, 12pt]{article}
\usepackage[T1]{fontenc}
\usepackage{babel}
\usepackage{ifsym}
\usepackage{geometry}
\usepackage{amsmath, amsthm, amssymb}
\usepackage{graphicx}
\usepackage{booktabs}
\usepackage{rotating}
\usepackage{setspace}
\usepackage{authblk}  
\usepackage{epigraph}
\usepackage{enumitem}   
\usepackage{url}
\usepackage[round]{natbib}
\usepackage{multirow}
\usepackage{xcolor}
\usepackage[toc,page]{appendix}

\usepackage{import}
\usepackage{xifthen}
\usepackage{pdfpages}
\usepackage{transparent}

\usepackage[colorlinks,
linkcolor=red,
anchorcolor=blue,
citecolor=blue
]{hyperref}

\newtheorem*{theorem*}{Theorem}

\newtheorem{Proposition}{Proposition}

\newtheorem*{Assumption*}{Assumption}
\newtheorem*{Definition*}{Definition}
\newtheorem{Obs}{Observation}

\makeatletter
\usepackage{babel}
\usepackage{bbm}
\usepackage{titlesec}
\titleformat*{\subsubsection}{\normalfont\fontsize{12}{17}\itshape}

\makeatother
\begin{document}

\title{Inefficient Peace or Preventive War?}


\date{}

\author{Liqun Liu\thanks{Email: liuliqunallen@gmail.com.} \; Tusi (\"{U}ndes) Wen\thanks{Email: tusi.undes.wen@gmail.com.}}
\maketitle
\begin{abstract}


We study a model of two-player bargaining game in the shadow of a preventive trade war that examines why states deliberately maintain trade barriers in the age of globalization. Globalization can induce substantial power shifts between states, which makes the threat of a preventive trade war salient. In this situation,  there may exist ``healthy'' levels of trade barriers that dampen the war incentives by reducing states' expected payoffs from such a war. Thus, we demonstrate that trade barriers can sometimes serve as breaks and cushions necessary to sustain inefficient yet peaceful economic cooperation between states. We assess the theoretical implications by examining the US-China trade relations since 1972.

	\vspace*{1cm}
			
			\noindent \textbf{Word Count:} 7480
			\vspace*{1cm}
\end{abstract}
 \thispagestyle{empty}
\clearpage
\setcounter{page}{1}
\epigraph{\itshape Under a system of perfectly free commerce, each country naturally devotes its capital and labour to such employments as are most beneficial to each. This pursuit of individual advantage is admirably connected with the universal good of the whole. 
}{--- David Ricardo, \textit{On the Principles of Political Economy and Taxation}}

\epigraph{\itshape And the truth is that from the 1930s up to Donald Trump, the U.S. government did, in fact, pursue a strategy of negotiating reductions in tariffs and other barriers to trade, in the belief that more trade would both foster economic growth and, by creating productive interdependence among nations, promote world peace.\footnotemark}{--- Paul Krugman}
\footnotetext{https://www.nytimes.com/2021/08/20/opinion/us-globalization-tariffs.html}

\section{Introduction}







Even at the height of globalization, trade barriers are everywhere. Over the past three decades,  although trade barriers have been declining, the average tariff rate has remained at non-negligible levels above 5\%.\footnote{Data source: https://data.worldbank.org/indicator/TM.TAX.MRCH.SM.AR.ZS} In recent years, protectionism was again on the rise, with several major economies adjusting their tariffs upward. Notably, the trade war between the United States and China since 2018 has witnessed sharply increased tariffs that risk an economic decoupling between the world's largest economies. Such an empirical pattern seems to contradict conventional trade theories:  if removing trade barriers promises a Pareto improvement for trading partners, why should trade barriers persist?

The prevailing scholarship usually treats persistent trade barriers as the economic outcome of the international system. For example, hegemonic powers may create stability that facilitates mutually beneficial international cooperation (e.g., \cite{krasner1976state, gowa1989bipolarity}). State preferences also matter for international relations (e.g., \cite{milner1997interests}, \cite{moravcsik1997taking}, and \cite{snyder1991myths}), and trading arrangements in particular (e.g., \cite{bailey1997institutional, mansfield2000free, mansfield2002democracies, milner2005move, oneal1999kantian}). However, this view seems to underappreciate the strategic value of trade barriers
in the age of globalization, where  economic power is  intertwined with geopolitical and military power. For example, economic convergence may structurally change the power relations between states \citep{kennedy2010rise, mearsheimer2019bound}; technological spillovers may structurally shape states' positions in the global supply chains \citep{gereffi2005governance, zhang2016innovation, morrison2008global}. This raises the possibility of substantial power shifts particularly between developing and developed states, which sows the seeds for a preventive war. 

Two examples illustrate this point. The United States and China renormalized trade relations in 1972 and gradually reduced trade barriers, which boosted commerce across the Pacific and fueled globalization in the post-Cold War era. At the same time, the underlying power relations were shifting in favor of China. This trend of economic cooperation continued before it was severely interrupted in 2018, when the US under President Donald Trump launched an unprecedented trade war to forestall China's rise. Likewise, the US-Japan  trade relations underwent a similar pattern from the end of World War II to 1987, when the Plaza Accord was signed. Thus, in the face of substantial power shifts, lower trade barriers seem to corroborate with a higher likelihood of costly trade wars.

In this study, we rationalize the persistence of trade barriers by focusing on one previously overlooked mechanism: they may help prevent preventive trade wars.  When substantial power shifts in favor of the developing state take place, they may give rise to the well-known commitment problem that motivates the developed state to launch a preventive war \citep{fearon1995rationalist, levy1987declining,powell2006war}.  We argue that when states compete for economic dominance, persistent trade barriers can sometimes be a peacemaker. This is because  the developed state has weaker incentives to fight  over a shrunk market owing to trade barriers than when it faces a frictionless market.





Motivated by the stylized facts, we ask: when can trade barriers prevent an otherwise inevitable preventive trade war? Under what conditions do states prefer such an ``inefficient peace'' to a more efficient trade environment carrying the risk of trade wars? 

To answer these questions, we develop a model of two-player bargaining in the shadow of a trade war. In our model, a rising power and a declining power engage in a full-information, infinite-horizon crisis bargaining over the terms of trade. The defining characteristic of a ``rising power'' is a power shift that increases its probability to win a trade war against the opponent in the future. Per the foundational approach \citep{fearon1995rationalist,powell2006war}, such a power shift creates the commitment problem, which incentivizes the declining power to attack the rising power today to forestall future disadvantaged bargaining positions. Our key innovation is introducing trade barriers as a strategic tool under the rising power's control. Trade barriers are costly to interstate economic activities, as they reduce trade flows below the efficient level. More importantly, trade barriers are often path-dependent. First, trade barriers between rival states coming out of wars are likely to persist. For example, US-China trade relations remain contentious after the Korean War and during much of the Cold War. Second, trade partners under globalization face domestic resistance against protectionism. It could be prohibitively costly for China to reverse its open-up policies and market-oriented reforms back to the days in the 1980s. We analyze the strategic value of maneuvering such trade barriers in preventing a preventive trade war. 

Formally, we examine the conditions under which
states can maintain peaceful trade relations by deliberately keeping trade barriers. For such an  ``inefficient peace'' to emerge as an equilibrium outcome, three conditions are essential. First, the power shift is sizable and thus war-provoking. If this condition fails, the rising state perceives no risk of preventive trade wars, so it will eliminate trade inefficiencies prior to the bargaining process. Second, trade barriers are sufficiently damaging to the value of future trade flows. This condition is very likely to hold if states have substantial unrealized gains from economic cooperation. When it fails, the declining state will launch a preventive war despite unrealized gains and/or trading opportunities in the future. Third, the rising power must prefer such an inefficient peace to an ``efficient war.'' That is, the rising power's peace payoff must be higher than its payoff of
a preventive war. If this condition fails, the rising power views the inefficient peace undesirable, and would rather fight a costly trade war over an enlarged market. 

Our formal analysis uncovers a novel mechanism in which states may achieve peaceful bargains in the face of substantial power shifts. Notably, maintaining trade barriers can be seen as the rising power's credible threat to ``burn money,'' which deters the declining power from launching a preventive war. Such a tactic may achieve peaceful bargains not because it directly signals the rising power's private resolve \citep{austen2000cheap}, but because it relaxes the commitment problem faced by the opposing state. Along this line, \cite{schram2021hassling} shows that the commitment problem can be resolved if 
a declining power initiates low-level conflicts to thwart the speed of power shifts. \cite{krainin2022preventive} shows that the commitment problem can be resolved if the capital market opens to rival states. Our model also resonates with the wisdom from ``economic peace'' \citep{gartzke2007capitalist,gartzke2010international,mousseau2013democratic,schneider2010capitalist}: a better prospect of future economic cooperation diminishes states' incentives to engage in costly conflicts, be it military or economic.

Our general model helps explain a wide range of trading arrangements between states. Specifically, it explains why certain levels of trade barriers and disputes can be paradoxically ``healthy,'' that is, they provide necessary breaks and cushions that prevent globalization from accelerating into trade wars. Without them, globalization may risk producing trade wars when it generates sufficiently large economic power shifts between states. In this regard, our theory complements the arguments that states may use gunboat diplomacy to shape trade blocs \citep{gowa1993power,cooley2019gun}. 


We assess the implication of the model by examining US-China trade relations. We argue that China's WTO accession in December 2001 is a watershed event altering the future trajectories of US-China trade relations. Although the removal of trade barriers improved efficiency and stimulated trade flows between the two countries, it also sowed the seeds for future power struggles that fomented through hyperglobalization into the present US-China trade war. Prior to China's WTO accession, there existed substantial trade barriers and corresponding trade disputes between these two countries. Although the relative economic power was shifting in favor of China, the cost did not justify the benefit of a preventive war, because the potentials of the Chinese market remained dormant. After China's WTO accession, the removal of trade barriers between the two ushered in the era of what some economists call hyperglobalization \citep{krugman2019globalization} with dramatically increased trade flows. The relative economic power further shifted in favor of China, but with trade barriers largely removed, the incentives of launching a preventive trade war are no longer constrained by concerns over the long-term implications of existing disputes.






\section{Model}
In the model, a rising power and a declining power are engaged in a standard full-information, infinite-horizon crisis bargaining game \`a la \cite{fearon1995rationalist} and  \cite{powell2006war}. Our departure from the foundational approach is the introduction of a trade barrier that reduces the value of the resource being disputed between states. We assume in the baseline model that the barrier stems from varied sources within the rising power (e.g., technological and institutional constraints), and can be eliminated by the rising power alone. In the supplementary appendix, we demonstrate that the same qualitative results carry through if the barrier can be removed only through the cooperation of two powers. 

The stage game starts with the rising power deciding whether to eliminate the trade barrier, whenever it exists. The elimination decision is irrevocable, and entails enduring policy effects. For example, as a precondition for its WTO accession, China must undergo market-oriented institutional reforms, which imposed path-dependent constraints on future policymakers. Following the elimination decision, the productivity of the market between two states is unleashed, which expands the size of resource being disputed now and forever. The downside of such an ``efficient'' elimination decision is that it may weaken the rising power's future bargaining position. 
For example, signing an international environment treaty often requires a new state to make noncontractible upfront investment in green technology (\citet{battaglini2016participation}). Conversely, if the barrier is allowed to exist, it will shrink the value of the resource currently being disputed. Since our goal is to illustrate the peace-keeping effect of maintaining the barrier, we abstract the cost of elimination away to create the best-case scenario that justifies this decision: if the rising power finds such a cheap elimination undesirable, it is unwilling to undertake more expensive elimination.

The stage game then moves on to the crisis bargaining. Each period, the rising power makes an ultimatum offer to the declining power. The offer is subject to the resource constraint within that period. The declining power can accept this offer, end the current-period bargaining peacefully, and move forward to the next period; or it can reject this offer and initiate a war. Here we interpret ``war'' as a trade war, which may take the form of embargoes, economic sanctions, or prohibitively high tariffs; we may also interpret ``war'' at its face value -- a military activity. Following the ``costly lottery'' approach \citep{fearon1995rationalist}, we  suppose that the war winner captures the entire future flow of resource. Per the conventional interpretation, this specification can be viewed as the winner of the trade war taking over the entire economic market. Alternatively, it can be interpreted as two states signing long-term contracts after the trade war, receiving a per-period division of resource commensurate with their present war capacities \citep{schram2021hassling}. The key feature is that, a war will permanently resolve the crisis {\it and} prevent the rising power from eliminating the barrier in the future. While it may seem too strong to assume  that a war will forever destroy the unrealized gains from trade between states, our substantive results remain valid if such a war can sufficiently reduce the likelihood of efficient trade activities in the future. We address this point in the supplementary appendix.  

Between period 1 and 2, a power shift occurs that   enables the rising power to win a trade war against the declining power more likely in the future. This power shift may come from the rising power's exogenous growing market power, its technological improvement, or its removal of institutional constraints that inhibit productivity (e.g., the ``openup'' policy in China since 1979). As is well-known (e.g., \citealt{fearon1995rationalist,powell2006war}), the declining power may strategically launch a preventive war in the face of the power shift to forestall disadvantaged positions in the future. 

The rising power's decision of \textit{not} eliminating the trade barrier takes into account its direct economic costs as well as its strategic value  in deterring a preventive war. If the rising power eliminates the trade barrier, and expects the declining power to respect a long-term peaceful deal, it faces a holdup problem: so long as the power shift is substantial and the war is cheap, the declining power would initiate a preventive trade war that is costly to both sides. Maintaining the trade barrier reduces the declining power's willingness to initiate a preventive war, because it
reduces the spoils of the war victor. The more inefficiencies the trade barrier induces, the larger its strategic value of preventing a preventive trade war, as the declining power has less to gain from such a war. Together, the strategic value of maintaining the trade barrier is positively correlated with its damage to the future value of the economic market, which must be traded off against the rising power's incentive to engage in a preventive war.

{\bf Setup:} Two states, \textbf{R} and \textbf{D}, are in a full-information, infinite-horizon crisis bargaining game. \textbf{R} is a rising power, whereas \textbf{D} is a declining power.  Prior to the start of the game, there is an trade barrier between two states. We assign $t\in \{1,2,...\}$ to denote periods. 

The period-$t$ stage game starts with \textbf{R} choosing whether to eliminate  the trade barrier, whenever it exists. If the barrier has been eliminated (either before or within this period), the current and future flow of resource achieves its full potential, whose per-period value is normalized to 1. If the barrier is not eliminated by \textbf{R}, then it reduces the current value of resource being disputed  to $h_{t-1}$, which is realized and observed by both states prior to their strategic actions in period $t$.  Generally, $\{h_t\}_{t\geq 0}$ is a sequence of  identically and independently distributed random variables with cumulative density function $F([0,1])$ and mean $\mathbb{E}[h_t]=\mu$. Set  the default level of the barrier as $h_0\in (0,1)$.  

After the elimination decision, the stage game moves to crisis bargaining. If no war has occurred by the start of period $t$, both states observe the resource to be disputed in period $t$, which we denote by $y_t$. $y_t=1$ if the barrier has been eliminated; otherwise, $y_t=h_{t-1}$. The bargaining begins with \textbf{R} offering $x_t\in [0,y_t]$ to \textbf{D}. If \textbf{D} accepts the offer, then \textbf{R} receives $y_t-x_t$ and  \textbf{D} receives $x_t$; after that, nature draws $h_t$ from its distribution $F$ whenever possible, and the game moves forward to period $t+1$. If \textbf{D} rejects the offer, then war occurs, and the game ends. The winner captures all current and future resources.  In period $t$,
\textbf{D} wins the war with probability $p_t\in [0,1]$. The war entails a fixed and one-time cost $c_R>0$ for \textbf{R} and $c_D>0$ for \textbf{D}, which is common knowledge. Set  $p_{1}>p_{2}=p_{3}=...=p\in [0,1]$ to capture the idea that \textbf{D} is a declining power. That is, an exogenous power shift between period 1 and 2 makes \textbf{D} less likely to win a trade war in the future.

{\bf Strategies:} Let $\tau_{t}$ indicate whether a war has occurred by the end of period $t$. Since $y_t$ summarizes information about both the existence of the barrier and the size of resource being divided in period $t$, we denote  the  history at the beginning of period $t$ by $\mathcal{H}_{t}=\cup_{s<t}(\tau_{s}, y_{s},x_{s})$.
\textbf{R}'s decision to eliminate the barrier  can be written as $C_t:\mathcal{H}_t \rightarrow \{0,1\}$, where 1 means ``remove'' and 0 means ``keep.'' \textbf{R}'s period-$t$ bargaining strategy maps  from the history and the elimination decision to an offering strategy, which can be written as $P_t:\mathcal{H}_t\times \{0,1\}\rightarrow [0,1]$. \textbf{D}'s period-$t$ bargaining strategy maps from the history, the elimination decision, and \textbf{R}'s period-$t$ offer $x_t$, to the decision of acceptance/rejection, which can be written as $A_t:\mathcal{H}_t\times \{0,1\}\times [0,1]\rightarrow \{\text{Accept},\; \text{Reject}\}$.  If war occurs in period $t$, we denote two states' strategies as $\emptyset$ for period $t+1$
onward. 

\textbf{Preferences:} Both \textbf{R} and \textbf{D} are risk-neutral, and discount future payoffs at $\delta\in (0,1)$. If war occurs in period $t$, then the present-value war payoff for \textbf{R} is $ \mathbb{E}(1-p_t)\sum_{t\geq 1}\delta^{t-1} y_t-c_R$. Likewise, the present-value war payoff for \textbf{D} is  $ \mathbb{E}p_t \sum_{t\geq 1}\delta^{t-1} y_t-c_D$. If \textbf{D} accepts the offer $x_t$ in period $t$, then \textbf{R} receives $y_t-x_t$ and \textbf{D} receives $x_t$, and the game moves on to period $t+1$. If no war occurs between period $i$ and $j$, then \textbf{R} and \textbf{D} receive present-value payoffs (evaluated at period $i$) $\sum_{t=i}^j \delta^{t-1}(y_t-x_t)$ and $\sum_{t=i}^j \delta^{t-1}x_t$.

\textbf{Equilibrium concepts.} We focus on the subgame perfect Nash Equilibria (SPNE), which require states' strategies to be sequentially rational. We follow \citet{schram2021hassling}'s taxonomy to classify the set of SPNEs as below:

\begin{Definition*}
A SPNE is a war equilibrium if $\tau_t\neq 0$ for some $t\geq 1$. 
\end{Definition*}
In this type of equilibrium, war occurs. 

\begin{Definition*}
A SPNE is an efficient peaceful equilibrium if $\tau_t = 0$ and $y_t=1$ for all $t$.  
\end{Definition*}
In this type of equilibrium, war never occurs. Moreover, \textbf{R} removes barriers in period 1, leading to an efficient economic outcome. 

\begin{Definition*}
A SPNE is an inefficient peaceful equilibrium if $\tau_t = 0$ for all $t$, and $y_t\neq 1$ for some $t\geq 1$.  
\end{Definition*}
In this type of equilibrium, war never occurs. Moreover, \textbf{R} does not eliminate barriers at least in period 1, leading to an inefficient economic outcome.

We focus on the set of parameters under which (1) an efficient peaceful equilibrium does not exist; (2) an inefficient peaceful equilibrium exists. Together, they suggest that certain trade barriers  may actually help prevent a preventive trade war. If an efficient peaceful equilibrium exists, then it is  likely to be focal. For example, states tend to form international organizations to facilitate international cooperation \citep{keohane2005after}. This situation essentially replicates the  well-known war-inefficiency results or the Coase theorem \citep{coase1960}. If the war equilibrium is the only type of equilibrium that exists, then our full-information setup alludes to the commitment problem as the cause of war, which is also well-received in the formal models of conflicts (e.g., \cite{fearon1995rationalist,powell2006war}).  Thus, the parameter restriction allows us to make a strong case: sometimes seemingly inefficient trading arrangements can substitute for a costlier preventive war.

Before proceeding, we briefly compare our game form with related models. Our setup augments the canonical preventive war model with a decision to remove the trade barrier each period. The flip side of removing the barrier, ``maintaining the barrier,''  resembles what is called  ``hassling'' in \cite{schram2021hassling}. Both strategies are costly, with the crucial distinction being that maintaining the barrier does not affect the power shift between states, whereas hassling does. That said, maintaining the trade barriers reduces the size of resource being divided between states; hence, it may dampen the declining power's incentive to trigger a preventive war. 
\section{Analysis}
\subsection{Equilibria}
To begin, note that if the game ever moves forward to period $t\geq 2$, it entails neither uncertainty nor power shift. This means that states would like to resolve crisis peacefully and efficiently, as war is costly to both sides (see also \cite{fearon1995rationalist}). Specifically, \textbf{R} eliminates the barrier as early as possible to minimize inefficiencies, resulting in $y_t=1$ for all $t\geq 2$. Then we can invoke the results in \citet[pp186]{powell2006war}: Each period, subgame perfection requires  \textbf{R}'s  offer to make \textbf{D} exactly indifferent between peace and war.


\begin{Proposition}\label{P1}
If and only if $c_D\geq \Bar{c}_D(p,p_1):=(\frac{p_1-\delta p}{1-\delta}-1)/(1-\delta)$,  an efficient peaceful equilibrium exists. 
\end{Proposition}
\begin{proof}
In the Appendix. 
\end{proof}
The result illustrates why the commitment problem can be a cause of (preventive) wars: the threat of a rising power forces a declining power to  strategically attack  to forestall a rearing power shift. Such a war can be averted only if the declining power \textbf{D}'s war cost is sufficiently high. Since this war-inefficiency result is well-known,  we impose the following parameter restriction to focus on the interesting case. 
\begin{Assumption*}
$c_D<\Bar{c}_D(p,p_1)$. 
\end{Assumption*}
This assumption ensures that states cannot simultaneously eliminate two sorts of inefficiencies that we mentioned above: once \textbf{R} eliminates the trade barrier in period 1,  \textbf{D} would launch a preventive war to forestall disadvantaged positions due to power shifts in the future.  Now, we provide conditions under which trade barriers may prevent costlier preventive wars. 

\begin{Proposition}\label{P2}
If and only if $c_D\geq \underline{c}_D(\mu,p,p_1):= [\frac{\delta}{1-\delta}(\mu p_1-p)-(1-p_1)h_0]/(1-\delta)$ and $c_D+c_R\geq \underline{C}(\mu,p_1,h_0):= (1-p_1)(1-h_0)-\frac{\delta}{1-\delta} p_1 (1-\mu)$, an inefficient peaceful equilibrium exists.
\end{Proposition}
\begin{proof}
In the Appendix. 
\end{proof}
To sustain inefficient peace as an equilibrium outcome, two conditions must hold. First, trade barriers must sufficiently  reduce the \textbf{D}'s war spoils to render its cost of a preventive war unjustified. If this condition fails (e.g., $\mu\approx 1, h_0 \approx 1$), then \textbf{D} would launch a preventive war despite unrealized gains from efficient trade. Second, even if maintaining the barrier can be effective in dampening \textbf{D}'s war incentive, it is
nonetheless costly to \textbf{R}. Thus, to deter \textbf{R} from eliminating today's economic inefficiencies to embrace a preventive war, such war must be expensive enough to \textbf{R}. 

Combining Propositions \ref{P1} and \ref{P2}, an inefficient peaceful equilibrium exists if the preventive war is sufficiently costly to both sides, but not so costly to eliminate the war-provoking commitment problem from the start. It is also useful to point out that the parameters supporting an inefficient peaceful equilibrium is nontrivial in the parameter space. This is because in the $c_D$--$c_R$ plane, the areas bordered by $\Bar{c}_D(p,p_1)>c_D\geq \underline{c}_D(\mu,p,p_1)$ and $c_D+c_R\geq \underline{C}(\mu,p_1,h_0)$ always intersect.

\subsection{Implications}

Next, we provide more details about the existence conditions of an inefficient peaceful equilibrium. We highlight Observation \ref{O3}, as it entails rich empirical implications. Throughout, we maintain the assumption $c_D<\Bar{c}_D(p,p_1)$.  

\begin{Obs}\label{O1}
All else equal, a larger power shift (smaller $p$) or a lower war cost (smaller $c_D$) makes a preventive war more likely. Specifically, if $c_D\leq \underline{c}_D(\mu,p,p_1)$, a preventive war is inevitable.
\end{Obs}
Observation \ref{O1} follows directly from  the functional form of $\underline{c}_D(\mu,p,p_1)$: both a larger power shift and a lower war cost  makes a preventive trade war more attractive to the declining power. 

\begin{Obs}\label{O2}
Suppose the aggregate war cost $c_D+c_R$ is low (relative to $\underline{C}(\mu, p_1,h_0)$). Then even if \textbf{R} may deter a preventive war (which happens if $c_D>\underline{c}_D(\mu,p,p_1)$), it nonetheless is willing to eliminate the inefficiencies and go to war. 
\end{Obs}
Observation \ref{O2} describes the strategic value of {\it not eliminating} the trade barrier to \textbf{R}. Just as a trade war, the trade barrier also erodes the value of resource to be divided between states. When \textbf{R} finds a trade war cheap, it may actually prefer eliminating the barrier and going to an ``efficient'' preventive trade war, in lieu of leaving the barrier intact and signing an inefficient peace agreement.

\begin{Obs}\label{O3}
All else equal, if the expected level of trade barriers become costlier (smaller $\mu$), an inefficient peaceful equilibrium is more likely to exist. 
\end{Obs}
Observation \ref{O3} combines the results of Proposition \ref{P1}-\ref{P2}. In situations where \textbf{D} declines slow enough ($p$ large), a preventive war is unlikely, and  states would like to sign an efficient long-term peaceful contract to resolve disputes now and forever. In situations where \textbf{R} rises fast enough and becomes a threat to \textbf{D}, leaving the trade barriers intact becomes a viable means to deter a preventive war. 
Specifically, a smaller $\mu$ makes it cheaper for \textbf{R} to ``buy off'' \textbf{D} before the power shift is realized, and it does not change \textbf{R}'s cost of peace. As such, a smaller $\mu$ unambiguously makes an inefficient peaceful equilibrium more plausible.

\begin{Obs}\label{O4}
If the default level of trade barriers become costlier (smaller $h_0$), we cannot make definitive statements about the existence of an inefficient peaceful equilibrium. 
\end{Obs}
Observation \ref{O4} is in the same vein as Observation \ref{O3}: a higher default level of barrier (lower $h_0$) reduces \textbf{R}'s cost to buy off \textbf{D}. The crucial difference is that, a higher default level of barrier also raises the inefficiencies in the period-1 bargaining, rendering an ``efficient'' preventive war (after the elimination of the barrier) more appealing to \textbf{R}. Thus, its overall effect on the plausibility of an inefficient peaceful equilibrium is ambiguous. 

\begin{figure}[htpb]
	\centering
	\def\svgwidth{0.8\columnwidth}
	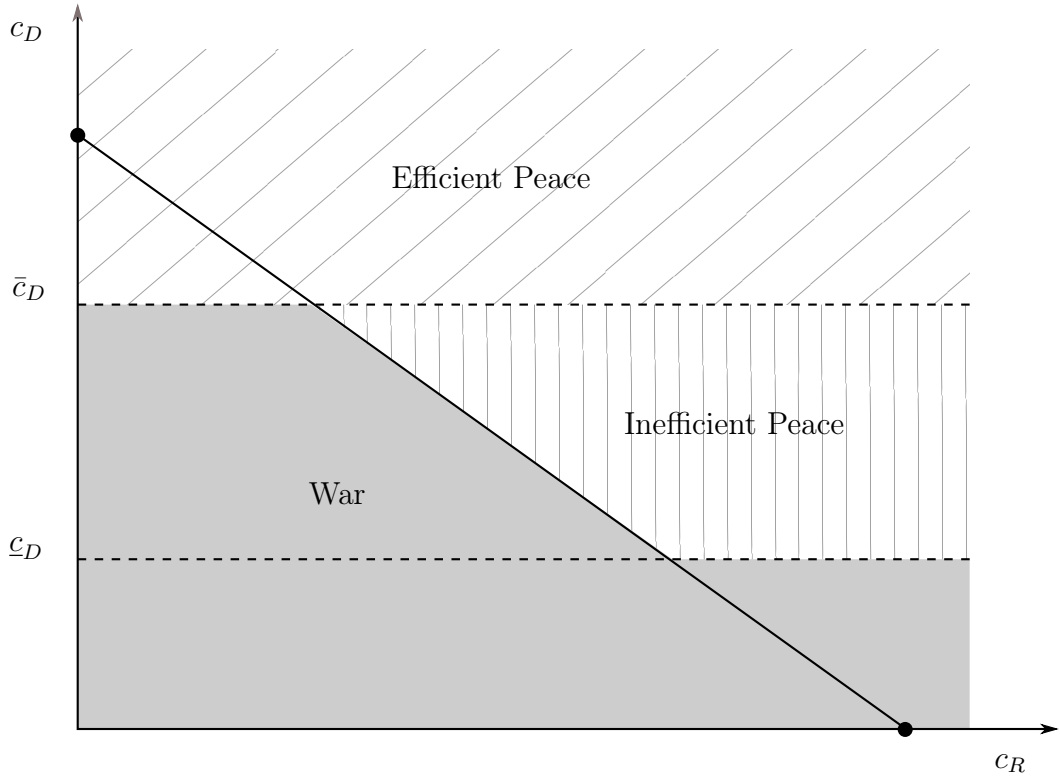

	\caption{Equilibria}
	\label{fig:equilibria}
\end{figure}

Figure~\ref{fig:equilibria} graphically summarizes our main results and their empirical implications. The two axes correspond to the costs of war for the rising power and declining power respectively. The expected future value of the disputed resources, the magnitude of the power shift, and costs of war jointly affect the likelihood of war and peace in the shadow of power shifts. The top horizontal dashed line represents the cost of war for the declining power $\overline{c}_D$ above which efficient peace can be sustained. When $c_D \geq \overline{c}_D$, the incentives for a preventive war disappear as \textbf{D}'s war cost completely overshadows any gains from a preventive war. Intuitively, peace can be maintained without any efficiency loss despite the war-inducing effect of power shifts when \textbf{D}'s cost of war is prohibitively high. The bottom horizontal dashed line represents the cost of war for the declining power $\underline{c}_D$ below which war becomes inevitable. When $c_D < \underline{c}_D$, the cost of war faced by the declining power is so low compared to the potential gains of winning the war that there is no way to prevent a preventive war. 

The most interesting scenarios arise in the region between these two horizontal dashed lines where $ \underline{c}_D < c_D \leq \overline{c}_D$. The slanted line represents a constraint on the joint costs of war for both \textbf{D} and \textbf{R} such that $c_D + c_R = \underline{C}(\mu, p_1,h_0)$ divides this region further into the left and right parts.
War arises in the left part where $c_D+c_R < \underline{C}(\mu, p_1,h_0)$. However, in the right part where $c_D+c_R \geq \underline{C}(\mu, p_1,h_0)$, it becomes possible to maintain peace at the cost of some efficiency loss. As the parameters of interests $\mu$, $p$, $c_D$, and $c_R$ change, boundaries between peace and war shift and the equilibrium outcomes change.
%

To situate the comparative statics in a real-world context, we  examine the US-China trade relations since 1972. 
\section{Case Studies}

In this section, we apply the model to address the puzzle raised at the beginning of this paper, i.e., why did the US and China become entangled in a trade war since 2018 after decades of economic cooperation through free trade? We view China's accession to the WTO in 2001 as a watershed event, as it has fundamentally altered the two countries' expectations of economic cooperation for the next two decades through hyperglobalization. 

After China joined the WTO,  both sides dramatically removed trade barriers. They not only managed to realize potential commercial gains by enlarging the size of the shared market, but also increased the stakes of future economic competition over the division of profits from global commerce. As the underlying distribution of power started to shift in favor of China, the US became more motivated to launch a preventive trade war to secure its future economic dominance. However, existing trade barriers served as breaks and cushions that prevent globalization from accelerating into a trade war. With the barriers removed and the potential gains from trade realized,  the expected value of future economic dominance outweighed the cost of a trade war. In the end, increased trade efficiency made possible by hyperglobalization paradoxically undermined the US's incentives to maintain peaceful economic cooperation with China. 



\subsection{The Pre-WTO US-China Trade Relations}

For the three decades following the founding of the People's Republic of China in 1949, US-China trade flows practically ground to a halt, since Washington and Beijing severed economic ties in a display of mutual hostility. As part of the Communist bloc and in the middle of the Cold War, China initially relied on Soviet support and devised its foreign policies under the Soviet tutelage, until the Sino-Soviet split in late 1950s created cleavages in the Communist camp.

As the Sino-Soviet relations deteriorated in the 1960s, both the US and China began to reassess the strategic value of the US-China relations in the context of the Cold War. It was against this historical backdrop that the US-China relations thawed with Nixon's ice-breaking meeting with Mao in 1972. Later, Deng carried on the momentum of improving US-China relations and in 1979 normalized the US-China trade relations, prompting an immediate growth of bilateral trade. Concomitant to the normalization of trade relations, China began the process of economic reform following the open-up policies. The market-oriented reform continued under Jiang. Many state-owned enterprises (SOE) were privatized in the 1990s to further boost the efficiency of an export-oriented economy. 

From the US-China rapprochement in 1972 to China's formal accession to the WTO in 2001, a relative power shift was already underway for roughly three decades fueled by persistent Chinese trade surplus toward the US. Despite the power shift, two countries maintained a largely amicable relationship with growing economic ties, as both sides brushed away ideological differences to jointly promote globalization. 

What made the US treat a rising China not as a threat but an economic ally despite apparent political incongruence during these three decades? In the context of US-China relations and the broader historical background, our model suggests that several factors jointly contributed to the peaceful economic coexistence: low expected market value due to high trade frictions, high cost of war, small magnitude of power shifts, and the intellectual backing by the modernization theory and the liberal ideal. 

\begin{figure}[htpb]
	\centering
	\includegraphics[width=0.8\textwidth]{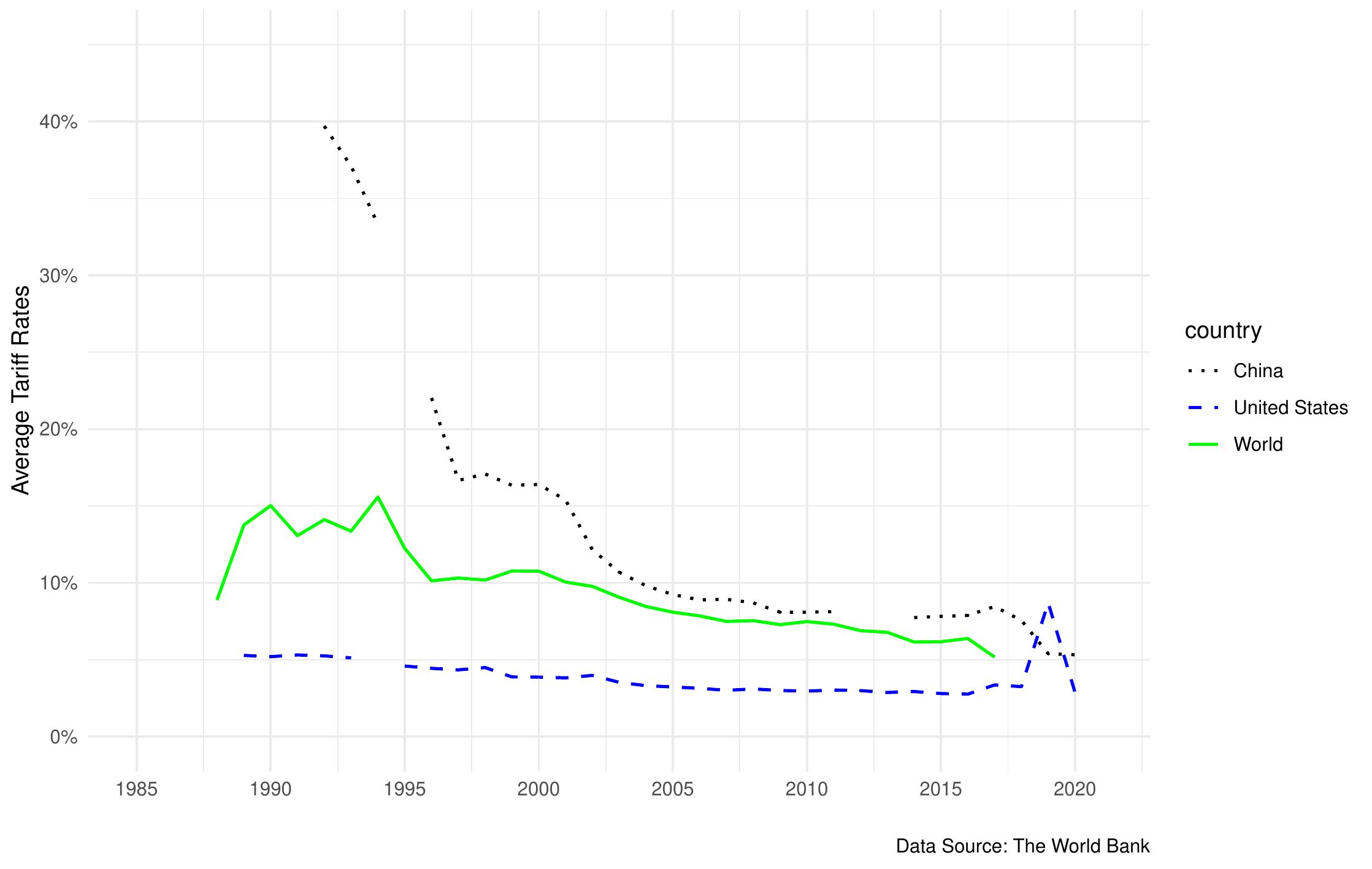}
	\caption{Average Tariff Rates for All Products, 1988-2020 (data source: the World Bank)}
	\label{fig:Tariffs}
\end{figure}
\begin{figure}[htpb]
	\centering
	\def\svgwidth{0.8\columnwidth}
\begingroup%
  \makeatletter%
  \providecommand\color[2][]{%
    \errmessage{(Inkscape) Color is used for the text in Inkscape, but the package 'color.sty' is not loaded}%
    \renewcommand\color[2][]{}%
  }%
  \providecommand\transparent[1]{%
    \errmessage{(Inkscape) Transparency is used (non-zero) for the text in Inkscape, but the package 'transparent.sty' is not loaded}%
    \renewcommand\transparent[1]{}%
  }%
  \providecommand\rotatebox[2]{#2}%
  \newcommand*\fsize{\dimexpr\f@size pt\relax}%
  \newcommand*\lineheight[1]{\fontsize{\fsize}{#1\fsize}\selectfont}%
  \ifx\svgwidth\undefined%
    \setlength{\unitlength}{373.81796935bp}%
    \ifx\svgscale\undefined%
      \relax%
    \else%
      \setlength{\unitlength}{\unitlength * \real{\svgscale}}%
    \fi%
  \else%
    \setlength{\unitlength}{\svgwidth}%
  \fi%
  \global\let\svgwidth\undefined%
  \global\let\svgscale\undefined%
  \makeatother%
  \begin{picture}(1,0.72614487)%
    \lineheight{1}%
    \setlength\tabcolsep{0pt}%
    \put(0,0){\includegraphics[width=\unitlength,page=1]{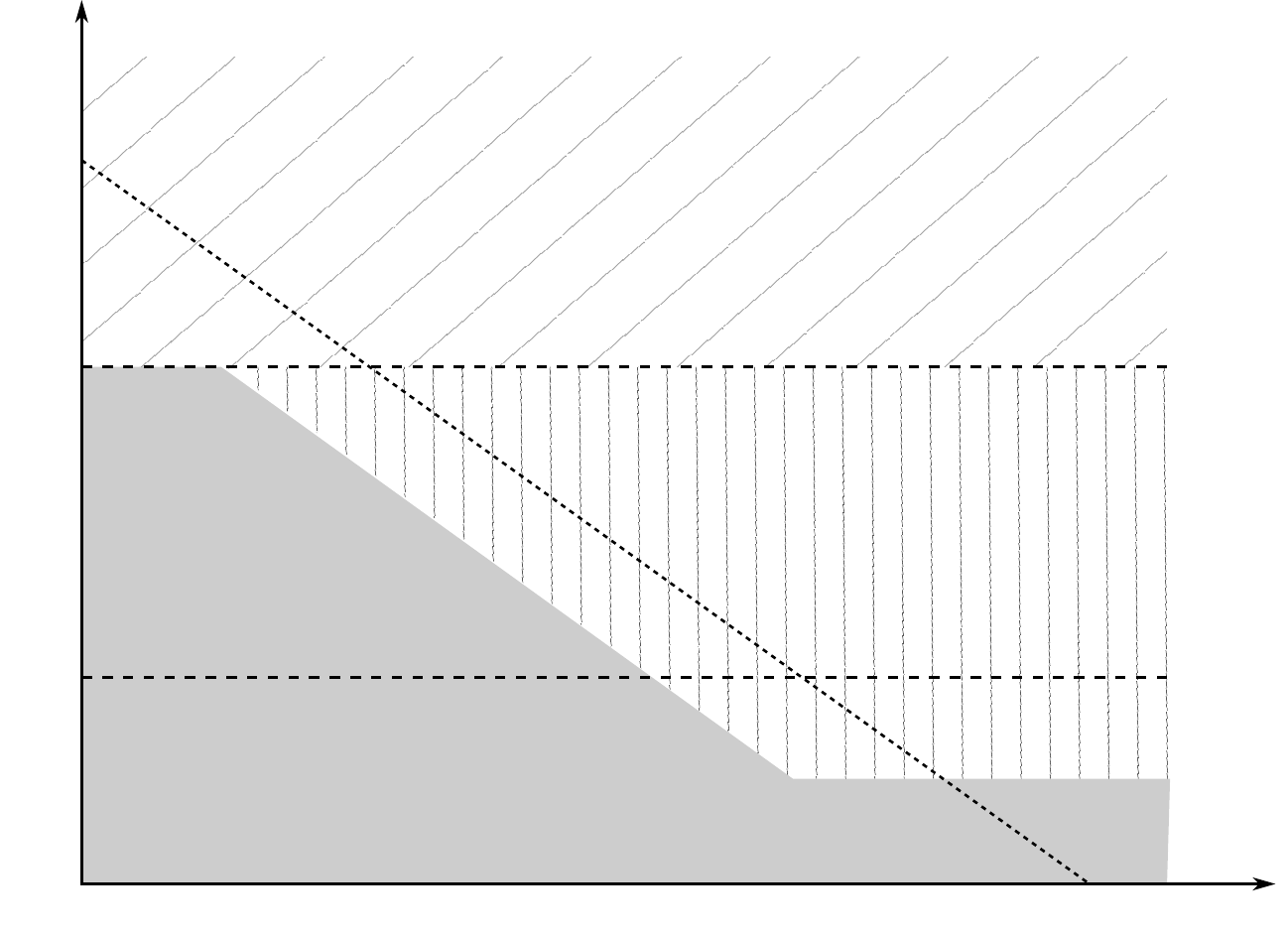}}%
    \put(0.92945211,0.00346274){\color[rgb]{0,0,0}\makebox(0,0)[lt]{\lineheight{1.25}\smash{\begin{tabular}[t]{l}$c_R$\end{tabular}}}}%
    \put(-0.00084487,0.69513495){\color[rgb]{0,0,0}\makebox(0,0)[lt]{\lineheight{1.25}\smash{\begin{tabular}[t]{l}$c_D$\end{tabular}}}}%
    \put(0.00166703,0.44978993){\color[rgb]{0,0,0}\makebox(0,0)[lt]{\lineheight{1.25}\smash{\begin{tabular}[t]{l}$\bar{c}_D$\end{tabular}}}}%
    \put(-0.00117036,0.20838395){\color[rgb]{0,0,0}\makebox(0,0)[lt]{\lineheight{1.25}\smash{\begin{tabular}[t]{l}$\underline{c}_D$\end{tabular}}}}%
    \put(0.2817169,0.25266446){\color[rgb]{0,0,0}\makebox(0,0)[lt]{\lineheight{1.25}\smash{\begin{tabular}[t]{l}War\end{tabular}}}}%
    \put(0.579641,0.31934279){\color[rgb]{0,0,0}\makebox(0,0)[lt]{\lineheight{1.25}\smash{\begin{tabular}[t]{l}Inefficient Peace\end{tabular}}}}%
    \put(0.35974461,0.5520071){\color[rgb]{0,0,0}\makebox(0,0)[lt]{\lineheight{1.25}\smash{\begin{tabular}[t]{l}Efficient Peace\end{tabular}}}}%
    \put(0,0){\includegraphics[width=\unitlength,page=2]{fig1.2.pdf}}%
  \end{picture}%
\endgroup%

	\caption{Lower expected future market value (smaller $\mu$) improves the likelihood of inefficient peace}
	\label{fig:equilibria2}
\end{figure}

First, the pre-existing trade barriers between the two countries provided necessary breaks and cushions that disincentivized the US from undermining a rising China. Over the three decades prior to China's WTO accession, although trade flows between the two countries steadily increased, various trade barriers remained in place. Figure~\ref{fig:Tariffs} illustrates the average tariff rates between the two countries from 1988 to 2020, which shows the elevated Chinese tariffs in the 1980s prior to China's GATT and WTO negotiations since 1986. As a result, the two countries had to continuously negotiate temporary trade agreements to resolve ongoing trade disputes. Such inefficiencies in bilateral trade translated into greater unrealized potential economic gains from free trade, which implies that the opportunity cost of forgoing continued economic cooperation was sufficiently large. Within the context of our model, this corresponds to a lower expected future market value $\mu$, which directly reduces the likelihood of a preventive war according to Observation~\ref{O3}. Graphically, a declining $\mu$ reduces $ \underline{C}(\mu, p_1,h_0)$ and $\underline{c}_D$, enlarging the zone of inefficient peace as shown in Figure~\ref{fig:equilibria2}.

Second, the US faced elevated cost of war against China in the shadow of the Cold War with the Soviet Union. The strategic value of China increased as its relations with the Soviet deteriorated during the Sino-Soviet split, which further increased the cost of any confrontation against China. Before the end of the Cold War in 1989, the primary adversary of the US was the Soviet Union. Normalizing trade relations with China brought both tangible and symbolic benefits to the US. The huge labor force and potential market of China were attractive for the US capital, and winning over China from the Soviet spheres of influence signified a symbolic victory of the US over the Soviet. From China's perspective, improving US-China relations enabled Mao and his successors to simultaneously pivot against the Soviet and the US, hedging geopolitical risks and increasing China's bargaining power against both. Normalizing trade relations with the US also provided tangible economic benefits by utilizing the huge Chinese labor force more efficiently. This explained why the US-China rapprochement initiated by the Nixon administration in 1972 was greeted by Mao despite the latter's hitherto staunch anti-US rhetoric. Thus, from 1972 and 1989, although the shadow of the US-Soviet Cold War loomed large, the risk of the US-China conflicts was substantially alleviated. In the context of our model, the higher cost for the US to engage in conflicts with China correspond to a greater $c_D$, thus increasing the likelihood that the equilibrium would fall into the peace zone. 

\begin{figure}[htpb]
	\centering
	\def\svgwidth{0.8\columnwidth}
\begingroup%
  \makeatletter%
  \providecommand\color[2][]{%
    \errmessage{(Inkscape) Color is used for the text in Inkscape, but the package 'color.sty' is not loaded}%
    \renewcommand\color[2][]{}%
  }%
  \providecommand\transparent[1]{%
    \errmessage{(Inkscape) Transparency is used (non-zero) for the text in Inkscape, but the package 'transparent.sty' is not loaded}%
    \renewcommand\transparent[1]{}%
  }%
  \providecommand\rotatebox[2]{#2}%
  \newcommand*\fsize{\dimexpr\f@size pt\relax}%
  \newcommand*\lineheight[1]{\fontsize{\fsize}{#1\fsize}\selectfont}%
  \ifx\svgwidth\undefined%
    \setlength{\unitlength}{376.09079842bp}%
    \ifx\svgscale\undefined%
      \relax%
    \else%
      \setlength{\unitlength}{\unitlength * \real{\svgscale}}%
    \fi%
  \else%
    \setlength{\unitlength}{\svgwidth}%
  \fi%
  \global\let\svgwidth\undefined%
  \global\let\svgscale\undefined%
  \makeatother%
  \begin{picture}(1,0.72175654)%
    \lineheight{1}%
    \setlength\tabcolsep{0pt}%
    \put(0,0){\includegraphics[width=\unitlength,page=1]{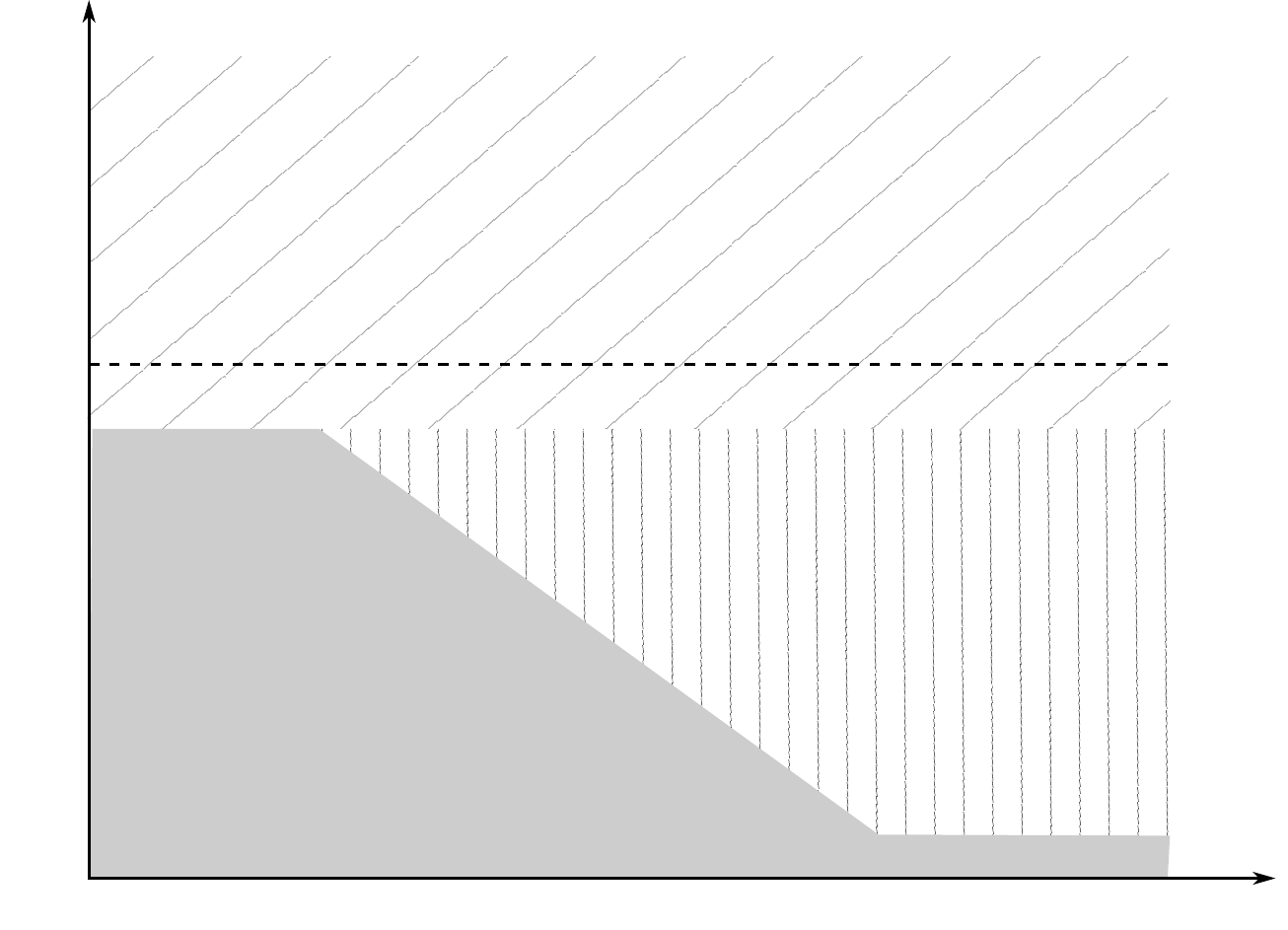}}%
    \put(0.92987845,0.00344181){\color[rgb]{0,0,0}\makebox(0,0)[lt]{\lineheight{1.25}\smash{\begin{tabular}[t]{l}$c_R$\end{tabular}}}}%
    \put(0.00520356,0.69093401){\color[rgb]{0,0,0}\makebox(0,0)[lt]{\lineheight{1.25}\smash{\begin{tabular}[t]{l}$c_D$\end{tabular}}}}%
    \put(0.00770029,0.44707172){\color[rgb]{0,0,0}\makebox(0,0)[lt]{\lineheight{1.25}\smash{\begin{tabular}[t]{l}$\bar{c}_D$\end{tabular}}}}%
    \put(-0.00116328,0.12856122){\color[rgb]{0,0,0}\makebox(0,0)[lt]{\lineheight{1.25}\smash{\begin{tabular}[t]{l}$\underline{c}_D$\end{tabular}}}}%
    \put(0.28605776,0.25113748){\color[rgb]{0,0,0}\makebox(0,0)[lt]{\lineheight{1.25}\smash{\begin{tabular}[t]{l}War\end{tabular}}}}%
    \put(0.58218135,0.31741285){\color[rgb]{0,0,0}\makebox(0,0)[lt]{\lineheight{1.25}\smash{\begin{tabular}[t]{l}Inefficient Peace\end{tabular}}}}%
    \put(0.36361386,0.54867107){\color[rgb]{0,0,0}\makebox(0,0)[lt]{\lineheight{1.25}\smash{\begin{tabular}[t]{l}Efficient Peace\end{tabular}}}}%
    \put(0,0){\includegraphics[width=\unitlength,page=2]{fig1.4.pdf}}%
  \end{picture}%
\endgroup%

	\caption{Smaller magnitude of power shift (larger $p$) is more likely to sustain peace}
	\label{fig:equilibria4}
\end{figure}

Third,  the magnitude of the power shift between the US and China remained moderate before China's WTO accession. For example, \citet{autor2016china} and \citet{hanson2019economic} find that China's economy had not started to hollow out the US manufacturing industry before 2001, which suggests that the distribution of power  between China and the US  were unlikely to change substantially back then. Since the future probably of winning a trade war against China, $p$, remained sufficiently high, the US was not afraid of China's rise, so it was unwilling to launch a preventive trade war. Graphically, as illustrated in Figure~\ref{fig:equilibria4}, a higher $p$ implies a higher likelihood of peace, as both $\underline{c}_D(\mu, p, p_1)$ and $\overline{c}_D(p,p_1)$ are decreasing in $p$.

Fourth, the modernization theory \citep{lipset1959some, przeworski1997modernization} and the liberal ideal \citep{mearsheimer2019bound} provided the intellectual backing for US rapprochement with China by mitigating the perceived political incongruence between the two countries despite a history of ideological animosity. According to the modernization theory, economic development is the precursor for democratic political reform. As an authoritarian country reaches a threshold of economic prosperity it will embark on the process of democratization. The liberal ideal provides a normative endorsement of global democratic promotion \citep{fukuyama1989end}. The modernization theory guided by the liberal ideal may have influenced the US decision-making, especially during the 1990s, to adopt a more optimistic view toward the democratic prospect of China and US-China relations. Within the context of our model, a democratic China is unlikely to engage in geopolitical contestation with the US. The perception that democratic countries resolve their disputes peacefully \citep{huntley1996kant, hegre2014democracy} implies that the opportunity cost of conflicts further increases, which improves the likelihood that the US-China trade relations would remain in the peace zone in Figure~\ref{fig:equilibria4}.


From the US-China rapprochement in 1972 to China's WTO accession in 2001, the three decades witnessed a gradual improvement of US-China relations and steady growth of trade flows between the two countries. The end of the Cold War in 1989 symbolized the victory of capitalism in the ideological confrontation and boosted the optimism of US leadership in embracing China as a partner of globalization. Prior to China's WTO accession, the pre-existing trade barriers remained in place and required periodic renegotiations to sustain trade flows between the two countries.  Paradoxically, such inefficiencies provided the necessary breaks and cushions that dampened the incentives for a preventive trade war, making possible a prolonged period of economic cooperation between otherwise unlikely partners.

\subsection{Post-WTO US-China Trade Relations}
After nearly five decades of economic cooperation, why would US-China trade relations hit a U-turn since the Trump administration launched the trade war in 2018? Our model suggests that China's WTO accession in 2001 marked a watershed event in the US-China relations, ushering in nearly two decades of explosive growth. The resulting rapid rise of China accelerated the power shift between the two countries. Just as the political and economic barriers paradoxically facilitated economic cooperation between the two unlikely partners in the middle of the Cold War, the removal of trade barriers with China's WTO accession paradoxically reduced the prospect of continued economic cooperation and increased the likelihood of a preventive trade war.

Before proceeding, we highlight the path-dependence or the ``stickiness'' of trade barriers, which is essential to our narratives. 
When applying for the WTO accession, China underwent market-oriented institutional reforms to conform to the standard and expectations consistent with WTO membership. Concomitantly, the reform-minded Chinese leadership since Deng promoted cohorts of well-educated technocrats to facilitate market-oriented institutional reforms. The resulting partial liberalization of the Chinese economy eventually qualified China for WTO membership. Meanwhile, such pro-market institutions were further reinforced through the vested interests of reformist cadres and business elites with party ties. 
The institutional path-dependence implies that China not only has strong incentives to embrace globalization for economic benefits, but also China cannot drastically reverse  its institutional reform.


Prior to China's accession to the WTO in 2001, trade flows between US and China were restricted by barriers erected on both sides. The full potentials from free trade were yet to be realized. With these trade barriers in place, the expected payoffs from future trade flows remained moderate. In other words, the expected annual market value $\mu$ fell short of its full potential of $1$. Therefore, although China had already been on its growth trajectory after the country's leadership adopted the open-up policies in 1978, the relative power shift between US and China did not overshadow the potential mutual gains from further expansion of trade relations. Referring to Figure~\ref{fig:equilibria}, the two countries found themselves in the inefficient peace zone. The inefficiency due to existing trade barriers promised greater potential gains from trade liberalization and thus imposed greater opportunity costs if the two sides were to turn hostile against each other. Removing trade barriers would thus unleash the efficiency gains and expand the shared market, which motivated China to apply for WTO accession and the US to eventually welcome China onboard.

Although some US policymakers today seemed to lament on their predecessors' decision of accepting China into the WTO, the decision was the result of lengthy deliberation and careful assessment for 15 years, and would not have been agreed upon had it not been deemed beneficial for both US and China. The reality of a relative power shift between the two countries had also been long recognized by US leadership. Mickey Kantor, the former US Trade Representative (1993-1996) and the Secretary of Commerce (1996-1997) under the Clinton Administration, recently pointed out when reflecting on the decision of approving China's WTO application: \textit{``I think we were generally correct that China was going to become the second largest power on the face of the earth and if the first largest power, the United States of America, did not reach out and begin to bring China into a system of liberal order and rule of law, then everyone was going to suffer...We knew everything would not go perfectly. President Clinton said it. President Bush and President Obama said it. No one had unrealistic expectations.''}\footnote{\textit{https://www.politico.com/news/2021/12/09/china-wto-20-years-524050}} Given the consensus of US decision-makers dated back to at least the Clinton administration that the relative power shift between US and China had been underway, the rise of China should not be regarded as a surprise. Considering the positive-sum nature of trade, why did a trade war break out after decades of voluntary economic cooperation that benefited both sides?

\begin{figure}[htpb]
	\centering
	\includegraphics[width=0.9\textwidth]{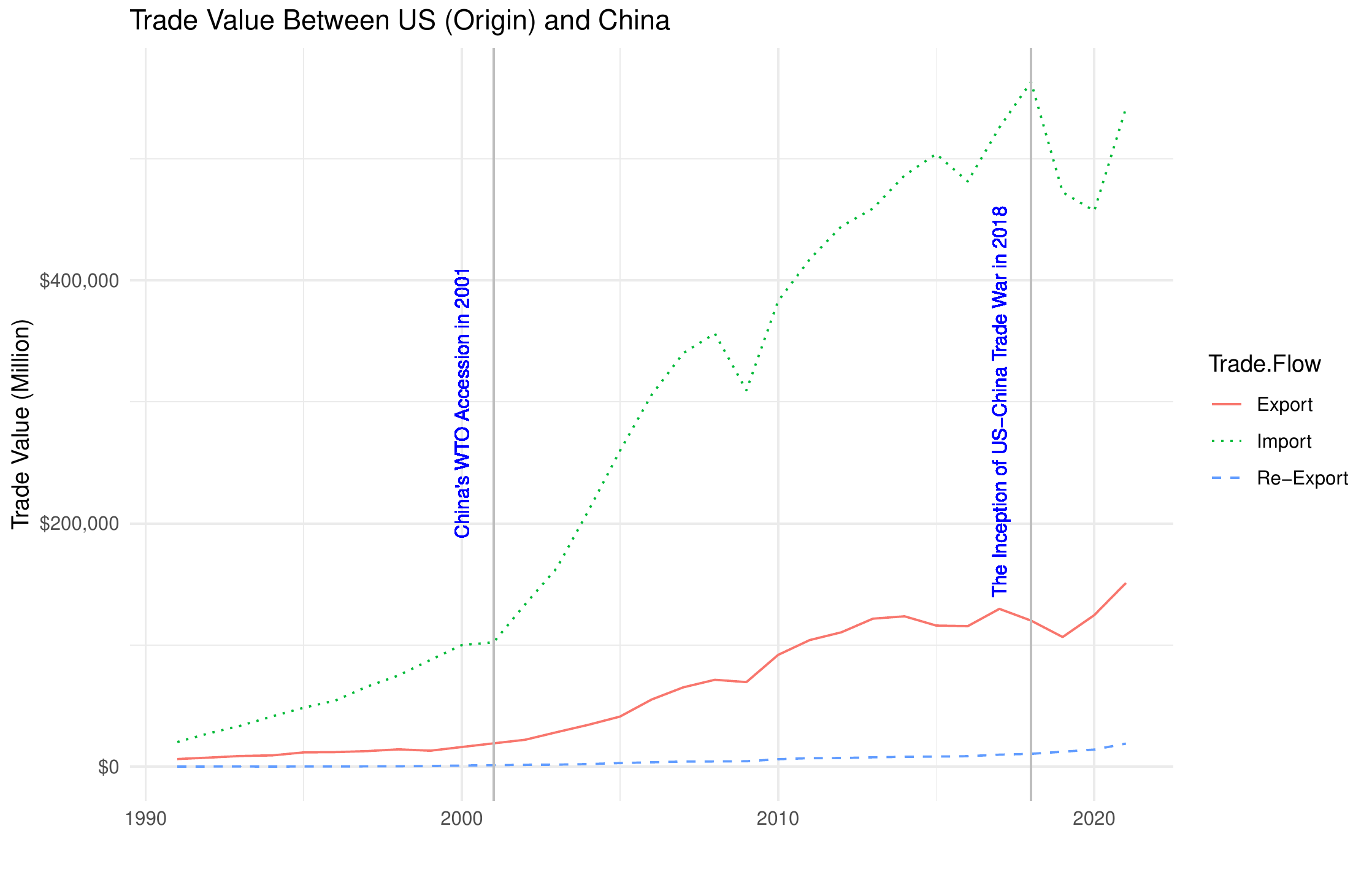}
	\caption{Value of Trade Flows Between US and China (data source: UN Comtrade)}
	\label{fig:TradeFlow}
\end{figure}

The surprise to all parties, however, is the magnitude of trade expansion and the speed of the resulting power shift after China joined WTO, which amounted to what some economists call hyperglobalization \citep{krugman2019globalization}. Figure~\ref{fig:TradeFlow} depicts the trajectories of trade flows between US and China from the early 1990s to 2021. Prior to its WTO accession, China had already posted consistent trade surpluses against the US, with widening gap between export and import every year. After China's WTO accession, although US export to China grew at a higher rate compared to the pre-WTO era, the import growth leaped forward, further enlarging the US trade deficit. The removal of pre-existing trade barriers after China's WTO accession and the consequent expansion of shared market and increased trade flows meant that both countries were poised to reap the full benefits of previously unrealized economic potentials of free trade. In the context of our model, this corresponds to the increase of expected annual market value $\mu$ toward the maximum level of $1$. Referring to Figure~\ref{fig:equilibria2}, the increase of $\mu$ contracts the area representing the inefficient peaceful equilibrium and expands the area representing (trade) war, making a preventive (trade) war more likely. In other words, the magnitude and swiftness of trade expansion over the two decades after China's WTO accession fundamentally altered the two countries' prospects and incentives of continued economic cooperation.


Furthermore, hyperglobalization increased the magnitude of a global power shift between US and China. On top of the growing US-China trade imbalance, China's success in accelerating scientific and technological advancement through technology transfers and domestic R\&D enabled Chinese companies to move upstream along the global supply chains, profiting from higher value-added manufacturing segments. The increasing global competitiveness of Chinese companies in a growing array of industries including high tech began to challenge the US dominance in crucial links of the global supply chains. The strategic investments and growing presence of Chinese companies in geographical areas that had been neglected by the US (such as Africa) alarmed the US for potentially losing grounds in the global geopolitical contestation. All of these factors jointly increase the prospective power shift as reflected in a diminished probability $p$ for the US to win a future conflict against China. Referring to Figure~\ref{fig:equilibria4}, a declining $p$ increases both $\underline{c}_D(\mu,p,p_1)$ and $\overline{c}_D(p,p_1)$, turning part of the inefficient peace zone into the war zone. 


Finally, the cost of conflict for the US $c_D$ declines, increasing the incentives for the US to launch a preventive trade war. Under the leadership of Xi Jinping, China began to pursue a more assertive political agenda that emphasized the country's rise as a global super power. Contrary to the prediction of the modernization theory, economic development through globalization did not manifest into democratization. Instead, China increasingly projected power on the global stage through grand strategies such as the ``Belt and Road Initiative.'' As US and its Western allies increasingly viewed China's rise as a threat to the existing global order, the political rhetoric toward China turned from skeptical cordiality to open hostility. From the perspective of the US leadership, the growing consensus that a rising China would challenge the West-dominated world order motivated the allies to cooperate with the US to forestall China's rise. Accordingly, the war cost borne by the US alone, $c_D$, declines. This moves the equilibrium downward along the y-axis, and increases the probability of war.

In sum, all aforementioned factors increased the incentives for the US to launch a preventive war against China in 2018. Since China's WTO accession, the economic potentials of trade liberalization had been realized ($\mu$ approaches $1$), making it more attractive for the US to seize the market. As it had become a widespread perception that a rising China was challenging the global order, the cost of trade war for the US declined. Together with an impending power shift accelerated by hyperglobalization, the benefits of a preventive trade war  overshadowed its cost.

More recently, the Covid-19 pandemic in 2019 imposed an exogenous shock that reduced existing trade flows, resulting in a lower $\mu$.  According to our theory, this change would alleviate states' propensities for a trade war. Specifically, the disruption to supply chains due to the pandemic not only increased trade frictions between US and China, but also increased the scope of peace. The Federal Reserve's stimulus response to the pandemic and the shortages due to supply chain disruptions created inflationary pressures in the United States, which further undercut the incentives to continue the trade war. So far, the Biden administration has softened its tones against China and lifted some of the Trump-era tariffs to battle inflation compounded by a potential recession. 
Therefore, we may expect the pandemic and the ensuing economic crisis to deescalate US-China conflicts. Nonetheless, the present truce between the US and China may prove temporary if the global trade can be revived and the power shift between US and China is to continue.

\section{Conclusion}

 In international politics, wars are costly means to end crises, as they impose a cost to all states. By the same token, trade barriers are costly for international economic activities, as they engender a dead weight loss for trading partners. We ask: if trade barriers are inefficient, why should they exist at all?  
    
We demonstrate with a formal model that the complete removal of trade barriers can be counterproductive, because doing so may trigger a preventive war between states.
When states expect a substantial power shift, a free trade environment may incentivize a declining power to launch a preventive trade war to assert its economic dominance in the future. In this situation, the rising power can strategically retain trade barriers or leave certain trade disputes unresolved to reduce a declining power's war payoff, thereby decreasing the likelihood of war.


Our results are sufficiently general to explain diverse conflict situations in the shadow of shifting power. Applying our model to studying the US-China trade relations leading up to the present trade war, we demonstrate the stabilizing effects of trade barriers and the logic behind the onset of trade wars. These insights challenge the received wisdom that globalization promotes peace by facilitating the economic interdependence between states. We argue that, perhaps paradoxically, sometimes it is the seemingly inefficient trade barriers that maintain the economic interdependence in the shadow of power shifts. 


\clearpage
\appendix
\begin{appendices}
\begin{proof}[Proof of Proposition \ref{P1}]
Efficiency requires $y_t=1$ for all $t\geq 1$. Let $x_1$ denote \textbf{R}'s  offer to \textbf{D} in period 1. Now, we invoke equation (2) in \cite{powell2006war}. By accepting the offer, \textbf{D} obtains an expected payoff of $x_1+\delta(\frac{p}{1-\delta}-c_D)$. By rejecting the offer, \textbf{D} obtains an expected payoff of $\frac{p_1}{1-\delta}-c_D$. Hence, \textbf{D} is willing to accept the offer whenever $x_1+\delta(\frac{p}{1-\delta}-c_D)\geq \frac{p_1}{1-\delta}-c_D$, which simplifies to $x_1\geq \frac{p_1-\delta p}{1-\delta}-(1-\delta) c_D$. Sequential rationality requires \textbf{R} to offer exactly $x_1^*=\frac{p_1-\delta p}{1-\delta}-(1-\delta) c_D$ to \textbf{D}. Such an offer is feasible if $x_1^*\leq 1$; that is, $(1-\delta) c_D\geq \frac{p_1-\delta p}{1-\delta}-1$. This shows necessity. To see the sufficiency part, we check if \textbf{R} may benefit by engaging in a war in period 1, since the crisis bargaining after period 2 must be peaceful. \textbf{R}'s war payoff is $\frac{1-p_1}{1-\delta}-c_R$, which is lower than its equilibrium payoff $\frac{1-p_1}{1-\delta}+c_D$. \end{proof}

\begin{proof}[Proof of Proposition \ref{P2}]
 In period 1, the size of resource being divided is $h_0$. Let $x_1$ denote  \textbf{R}'s  offer to \textbf{D} in period 1. By accepting the offer, \textbf{D} obtains an expected payoff of $x_1+\delta (\frac{p}{1-\delta}-c_D)$. By rejecting the offer, \textbf{D} obtains an expected payoff of $p_1 h_0+\frac{\delta}{1-\delta} p_1 \mu  -c_D$. Hence, \textbf{D} is willing to accept the offer whenever $x_1+\delta(\frac{p}{1-\delta}-c_D)\geq p_1 h_0+\frac{\delta}{1-\delta} p_1 \mu  -c_D$, which simplifies to $x_1\geq p_1 h_0-(1-\delta)c_D+\frac{\delta}{1-\delta}(\mu p_1-p)$. Sequential rationality requires \textbf{R} to offer exactly $x_1^*= p_1 h_0-(1-\delta)c_D+\frac{\delta}{1-\delta}(\mu p_1-p)$ to \textbf{D} whenever possible. Such an offer is feasible if $x_1^*\leq h_0$; that is, $c_D\geq \underline{c}_D:=[\frac{\delta}{1-\delta}(\mu p_1-p)-(1-p_1)h_0]/(1-\delta)$. 

Turn to the sufficiency part. It suffices to check if \textbf{R} may gain by removing barriers in period 1, followed by a preventive war, as this is \textbf{R}'s best deviation. \textbf{R}'s payoff in this case is $\frac{1-p_1}{1-\delta}-c_R$. Its equilibrium payoff is $h_0+\frac{\delta}{1-\delta}-[p_1 h_0+\frac{\delta}{1-\delta} p_1 \mu]  +c_D$, which simplifies to  $h_0(1-p_1)+\frac{\delta}{1-\delta} (1-\mu p_1)+c_D$. Hence, \textbf{R} does not want to deviate if $c_D+c_R\geq (1-p_1)(1-h_0)-\frac{\delta}{1-\delta} p_1 (1-\mu)$. 
\end{proof}
\end{appendices}
\clearpage
\renewcommand
\refname{Reference}
	\bibliographystyle{plainnat}
\bibliography{new.bib}
\end{document}


\title{Supplementary Appendix to ``Inefficient Peace or Preventive War''}

\author{}
\maketitle

The baseline model makes several  assumptions about trade barriers. First, they can be eliminated by the rising power alone. Second, after a full-scale preventive war, states can no longer eliminate trade barriers. Third, the existence of trade barriers does not affect the relative power between states. Here, we relax these assumptions to assess the model robustness. 

\section{When removing barriers requires cooperative efforts}
In the baseline model, trade barriers are modeled as the technological or institutional constraints faced by the rising power. While this notion of barriers applies to many settings, there are cases in which states must cooperate to eliminate the barriers. For example, establishing the most favored nation (MFN) treatment often requires each side to grant lower tariffs and quota to the other side. 

To incorporate such a concern, we modify the ``elimination decision'' of the stage game in the baseline model.  Instead of \textbf{R} choosing whether to eliminate the barrier, the modified period-$t$ stage game starts with \textbf{R} and \textbf{D} simultaneously choosing between whether to eliminate the trade barrier or not, whenever possible. Elimination requires both \textbf{R} and \textbf{D} to say ``Yes''; if either side says ``No,'' the barrier persists. Other game specifications remain the same as the baseline. 

We accordingly modify players' strategy set.  Let $C_t=(C_t^R,C_t^D)$ be states' decision to eliminate the barrier in period $t$. Denote the history at the beginning of period $t$ by $\mathcal{H}_t=\cup_{s<t}(C_s,y_s,x_s)$. \textbf{R} and \textbf{D}'s decisions to eliminate the barrier can be written as $C^i_t:\mathcal{H}_t\rightarrow \{0,1\}$, where 1 means ``Yes'' and 0 means ``No,'' where $i\in \{\textbf{R},\textbf{D}\}$. \textbf{R}'s period-$t$ bargaining strategy maps  from the history and the elimination decision to an offering strategy, which can be written as $P_t:\mathcal{H}_t\times \{0,1\}^2\rightarrow [0,1]$. \textbf{D}'s period-$t$ bargaining strategy maps from the history, the elimination decision, and \textbf{R}'s period-$t$ offer $x_t$, to the decision of acceptance/rejection, which can be written as $A_t:\mathcal{H}_t\times \{0,1\}^2\times [0,1]\rightarrow \{\text{Accept},\; \text{Reject}\}$.  If war occurs in period $t$, we denote two states' strategies as $\emptyset$ for period $t+1$
onward.  The solution concept is SPNE. The payoffs and the equilibrium taxonomy are the same with the baseline model. 

\begin{Proposition*}
(1) If and only if $c_D\geq \Bar{c}_D(p,p_1)$, where $\Bar{c}_D(p,p_1)$ is given by Proposition 1, then an efficient peaceful equilibrium exists. 
(2) If and only if $c_D\geq \underline{c}_D(\mu,p,p_1)$,  and $c_D+c_R\geq \underline{C}(\mu,p_1,h_0)$, where $\underline{c}_D(\mu,p,p_1), \underline{C}(\mu,p_1, h_0)$ are given by Proposition 2, an inefficient peaceful equilibrium exists. 
\end{Proposition*}
\begin{proof}
Part (1) and the necessity direction of Part (2) follows directly from the argument in Proposition 1. It suffices to construct a strategy profile that respects subgame perfection. Consider the following strategy profile: in the first stage, \textbf{R} says ``No'' whereas \textbf{D} says ``Yes'' to the elimination decision; \textbf{R} proposes $x_1=y_1p-c$ to \textbf{D}; \textbf{D} accepts any offer no less than $x_1$ \textit{and} (No, Yes) occurs on path, and fights otherwise. Each state chooses not to fight if (No, Yes) occurs on path, and fight otherwise. Starting from period 2 onward, each state says ``Yes'' to the elimination decision whenever possible; \textbf{R} proposes $x_t=y_tp-c$ to \textbf{D}; \textbf{D} accepts any offer no less than $x_t$ \textit{and} (Yes, Yes), and fight otherwise.  Since \textbf{R} and \textbf{D}'s payoffs are identical with the payoffs specified in Proposition 1, the conditions supporting each type of equilibrium are the same with Proposition 1. 
\end{proof}
This proposition shows that whenever one type of equilibrium exists in which \textbf{R} can eliminate the barrier alone, the same parameter restriction supports this type of equilibrium  when  \textbf{R} and \textbf{D} must cooperate to eliminate the barrier. Substantively, whenever \textbf{R} can veto the elimination decision, it can strategically maneuver the decision to achieve peaceful bargains. 

\section{Postwar cooperation}
In the baseline model, we assume that states after war never renormalize their trade relationship, and leave the economic inefficiencies due to trade barriers forever. We relax this assumption here. 

We suppose that after a costly war, the value of per-period resource $y_t$ achieves its full potential $1$ from then on with probability $\rho\in (0,1)$; with probability $1-\rho$, it stays at the level $h_{t-1}$ as specified in the baseline model. The parameter $\rho$ captures the possibility that states may renormalize their trade relationship due to diplomatic efforts; it may also capture the possibility that the victor of the trade war may benefit excessively from a shrunk monopolized market. Holding other assumptions intact, we arrive at the following Proposition: 

\begin{Proposition*}
(1) If and only if $c_D\geq \Bar{c}_D(p,p_1)$, where $\Bar{c}_D(p,p_1)$ is given by Proposition 1, then an efficient peaceful equilibrium exists. 
(2) If and only if $c_D\geq \underline{c}_D(\mu,p,p_1)$,  and $c_D+c_R\geq \underline{C}(\mu,p_1,h_0)$, where $\underline{c}_D(\mu,p,p_1):=[\frac{\delta}{1-\delta}(p_1\frac{(1-\rho)(1-\delta)\mu+\rho}{1-(1-\rho)\delta}-p)-(1-p_1)h_0]/(1-\delta)$ , an inefficient peaceful equilibrium exists. 
\end{Proposition*}
\begin{proof}
It suffices to compute the mean value of the postwar market, which we denote by $x$. By the recursive structure, $x=\rho\cdot 1+(1-\rho)[(1-\delta)\mu+\delta x]$, which gives $x=\frac{(1-\rho)(1-\delta)\mu+\rho}{1-(1-\rho)\delta}$. The result follows by substituting $x$ into $\mu$ in Proposition 1 and 2 in the main text. 
\end{proof}

Observe that $x=1$ if $\rho=1$ and $x=\mu$ if $\rho=0$; also $x$ is increasing and continuous in $\rho$. Thus, our baseline model can be seen as a special case of the general setup. We immediate conclude that if a costly war sufficiently dampens states' scope of future economic cooperation ($\rho\approx 0$), then our substantive results in the baseline model carry through. More generally, holding fixed parameters $(p,p_1, \delta,h_0)$, if there exists a neighborhood around $\mu$ that supports one type of equilibrium in the baseline model, then this type of equilibrium 
can also be supported by a neighborhood around the tuple $(\mu,\rho=0)$ in the extended model (by the continuity argument). 

\section{When barriers affect relative power}
In the baseline model, the existence of trade barriers  does not affect the probability that each state may win a war. We relax this assumption here. 

In the absence of trade barriers, we suppose that the probability that \textbf{R} wins a war in period $t$ is $p_1>p_2=p_3=...=p$, just as the baseline model. When trade barriers exist, the probability that \textbf{R} wins a war in period $t$ is $\theta p_1>\theta p_2=\theta p_3=...=\theta p$. Here $\theta$ measures to what extent the power relationship between states can be altered by trade barriers. We suppose $\theta\geq \frac{\mu+p-1}{\mu p}$ so that \textbf{R} embraces free trade after the power shift is realized. This specification rules out the uninteresting possibility that the rising power may deliberately use trade barriers to create a military advantage (for an extreme example, consider $\theta\approx 0$). We now redo the propositions in the main model.

\begin{Proposition}
If and only if $c_D\geq \Bar{c}_D(p,p_1):=[\frac{p_1-\delta p}{1-\delta}-1]/(1-\delta)$,  an efficient peaceful equilibrium exists. 
\end{Proposition}
\begin{proof}
Efficiency requires $y_t=1$ for all $t\geq 1$, which means that both states face exactly the same strategic situations as in the baseline model. This means that Proposition 1 fully carry through.  \end{proof}
\begin{Proposition}\label{P2}
If and only if $c_D\geq \underline{c}_D(\mu,p,p_1,\theta ):= [\frac{\delta}{1-\delta}(\mu \theta p_1-p)-(1-\theta p_1)h_0]/(1-\delta)$ and $c_D+c_R\geq \{\underline{C}(\mu,p_1,\theta,h_0):= 1-p_1-[(1-\delta)h_0(1- \theta p_1)+\delta (1-\mu \theta p_1)]\}/(1-\delta)$, an inefficient peaceful equilibrium exists.
\end{Proposition}
\begin{proof}
 In period 1, the size of resource being divided is $h_0$. Let $x_1$ denote  \textbf{R}'s  offer to \textbf{D} in period 1. By accepting the offer, \textbf{D} obtains an expected payoff of $x_1+\delta(\frac{p}{1-\delta}-c_D)$. By rejecting the offer, \textbf{D} obtains an expected payoff of $\theta p_1[h_0+\frac{\delta}{1-\delta}\mu]  -c_D$. Hence, \textbf{D} is willing to accept the offer whenever $x_1+\delta(\frac{p}{1-\delta}-c_D)\geq \theta p_1[h_0+\frac{\delta}{1-\delta}\mu]  -c_D$, which simplifies to $x_1\geq \theta p_1 h_0-(1-\delta)c_D+\frac{\delta}{1-\delta}(\mu \theta p_1-p)$. Sequential rationality requires \textbf{R} to offer exactly $x_1^*= \theta p_1 h_0-(1-\delta)c_D+\frac{\delta}{1-\delta}(\mu \theta p_1-p)$ to \textbf{D} whenever possible. Such an offer is feasible if $x_1^*\leq h_0$; that is, $c_D\geq \underline{c}_D:=[\frac{\delta}{1-\delta}(\theta\mu p_1-p)-(1-\theta p_1)h_0]/(1-\delta)$. 

Turn to the sufficiency part. It suffices to check if \textbf{R} may gain by removing barriers in period 1, followed by a preventive war, as this is \textbf{R}'s best deviation. \textbf{R}'s payoff in this case is $\frac{1-p_1}{1-\delta}-c_R$. Its equilibrium payoff is $h_0+\frac{\delta}{1-\delta}-\theta p_1[h_0+\frac{\delta}{1-\delta} \mu]  +c_D$, which simplifies to  $h_0(1- \theta p_1)+\frac{\delta}{1-\delta}(1-\mu \theta p_1)+c_D$. Hence, \textbf{R} does not want to deviate if $c_D+c_R\geq \underline{C}(\mu,p_1,\theta,h_0):= \{1-p_1-[(1-\delta)h_0(1- \theta p_1)+\delta (1-\mu \theta p_1)]\}/(1-\delta)$. 
\end{proof}

So, the results in the baseline model can be viewed as the special case of the propositions above, where $\theta$ is set to 1. Because we have assigned the same winning probability in the event of free trade as in the baseline model,  $\theta$ does not enter the conditions that determine when an efficient peaceful equilibrium is possible. However, $\theta$ may encourage or discourage the declining power \textbf{D} from launching a preventive war after the rising power \textbf{R} maintains trade barriers in period 1. 
Since both $\underline{c}_D$  and $\underline{C}$ are increasing in $\theta$, we conclude that if trade barriers benefit the declining power militarily ($\theta>1$), then the rising power \textbf{R} will find it harder to appease \textbf{D} compared with the case where such an advantage is absent ($\theta=1$). This makes an inefficient peace less likely.